\def\year{2022}\relax
\documentclass[letterpaper]{article} 
\usepackage{aaai23}  
\usepackage{times}  
\usepackage{helvet}  
\usepackage{courier}  
\usepackage[hyphens]{url}  
\usepackage{graphicx} 
\usepackage{natbib}  
\usepackage{caption} 
\DeclareCaptionStyle{ruled}{labelfont=normalfont,labelsep=colon,strut=off} 
\usepackage{amssymb}
\usepackage{amsmath}
\usepackage{color}
\usepackage{verbatim}
\usepackage{algorithm}
\usepackage{algorithmic}

\usepackage{newfloat}
\usepackage{listings}
\floatstyle{ruled}
\newfloat{listing}{tb}{lst}{}
\floatname{listing}{Listing}
\title{Incentive-boosted Federated Crowdsourcing}

\author{
   Xiangping Kang\textsuperscript{\rm 1},
    Guoxian Yu\textsuperscript{\rm 1,2,}\thanks{Corresponding author: Guoxian Yu.},
    Jun Wang\textsuperscript{\rm 2},
    Wei Guo\textsuperscript{\rm 2},
    Carlotta Domeniconi\textsuperscript{\rm 3},
    Jinglin Zhang\textsuperscript{\rm 4}
}
\affiliations{
    \textsuperscript{\rm 1}School of Software, Shandong University, Jinan, China\\
    \textsuperscript{\rm 2}SDU-NTU Joint Centre for AI Research, Shandong University, Jinan, China\\
    \textsuperscript{\rm 3}Department of Computer Science, George Mason University, Fairfax, VA, USA\\
    \textsuperscript{\rm 4}School of Control Science and Engineering, Shandong University, Jinan, China\\
    x.p.kang@mail.sdu.edu.cn, \{gxyu, kingjun, guowei, jinglin.zhang\}@sdu.edu.cn, carlotta@cs.gmu.edu
}
\begin{document}

\maketitle

\begin{abstract}
Crowdsourcing is a favorable computing paradigm for processing computer-hard tasks by harnessing human intelligence. However, generic crowdsourcing systems may lead to privacy-leakage through the sharing of worker data. To tackle this problem, we propose a novel approach, called \textbf{iFedCrowd} (\underline{i}ncentive-boosted \underline{Fed}erated \underline{Crowd}sourcing), to manage the privacy and quality of crowdsourcing projects. iFedCrowd allows participants to locally process sensitive data and only upload encrypted training models, and then aggregates the model parameters to build a shared server model to protect data privacy. To motivate workers to build a high-quality global model in an efficacy way, we introduce an incentive mechanism that encourages workers to constantly collect fresh data to train accurate client models and boosts the global model training. We model the incentive-based interaction between the crowdsourcing platform and participating workers as a Stackelberg game, in which each side maximizes its own profit. We derive the Nash Equilibrium of the game to find the optimal solutions for the two sides. Experimental results confirm that iFedCrowd can complete secure crowdsourcing projects with high quality and efficiency.

\end{abstract}

\section{Introduction}
Crowdsourcing becomes increasingly popular in recent decades which coordinates the Internet workers to do micro-tasks so as to solve computer-hard problems (e.g., image annotation, answering database-hard queries~\cite{fan2015icrowd, tong2020spatial, kang2021crowdsourcing}). Moreover, the development of crowdsourcing platforms, such as Amazon Mechanical Turk\footnote{https://www.mturk.com/}, CrowdFlower\footnote{http://www.crowdflower.com/} and  Baidu Test\footnote{https://test.baidu.com/}, makes it  more convenient to get crowdsourced data by recruiting broad workers. However, prior researches have found that the submitted data can reveal crowd workers' private information, such as locations, vocal prints, face images and even business secrets~\cite{tong2020federated,zhao2021crowdsensing}.  With the increasing concerns and regulations on data security and personal privacy, 
data privacy in crowdsourcing is getting more and more vital. State-of-the-art protection techniques usually achieve privacy preservation through injecting imprecision, such as cloaking~\cite{pournajaf2014spatial, zhai2019towards}, inaccuracy (e.g., local differential privacy~\cite{wang2016differential, wang2018geographic}) to perturb crowd workers' sensitive information~\cite{wang2020federated}. Nevertheless, these techniques would inevitably lead to quality-loss crowdsourcing as they need to modify the original data.

To solve the above challenge, the federated learning (FL) paradigm is proposed ~\cite{mcmahan2017communication}. FL enables distributed computing nodes to collaboratively train models without exposing their sensitive data, thus realizing privacy-preserving model training with little loss (or even no loss) of model performance~\cite{wang2020federated,yang2019federated}. The crowdsourcing system typically outsources data collection tasks to Internet workers and then aggregates and analyzes the sensing data~\cite{capponi2019survey, gummidi2019survey,tu2020crowdwt,yu2020active}. Nevertheless, the centralized platform is generally untrusted and may leak workers' private information.  With the prevalence of mobile devices with increasing computation power (e.g., laptops and cell phones) and advanced network infrastructure (e.g., 5G),  we can directly outsource data \textit{processing} tasks instead of data collecting tasks to participants within the FL framework. Consequently, the collected data that involves private information can be kept locally without being exposed to other workers and the server. The significance in the lens of federated crowdsourcing has been well recognized~\cite{tong2020federated, wang2020federated, pandey2020crowdsourcing}. 

\begin{figure*}[t]
    \centering
    \includegraphics[width=16cm]{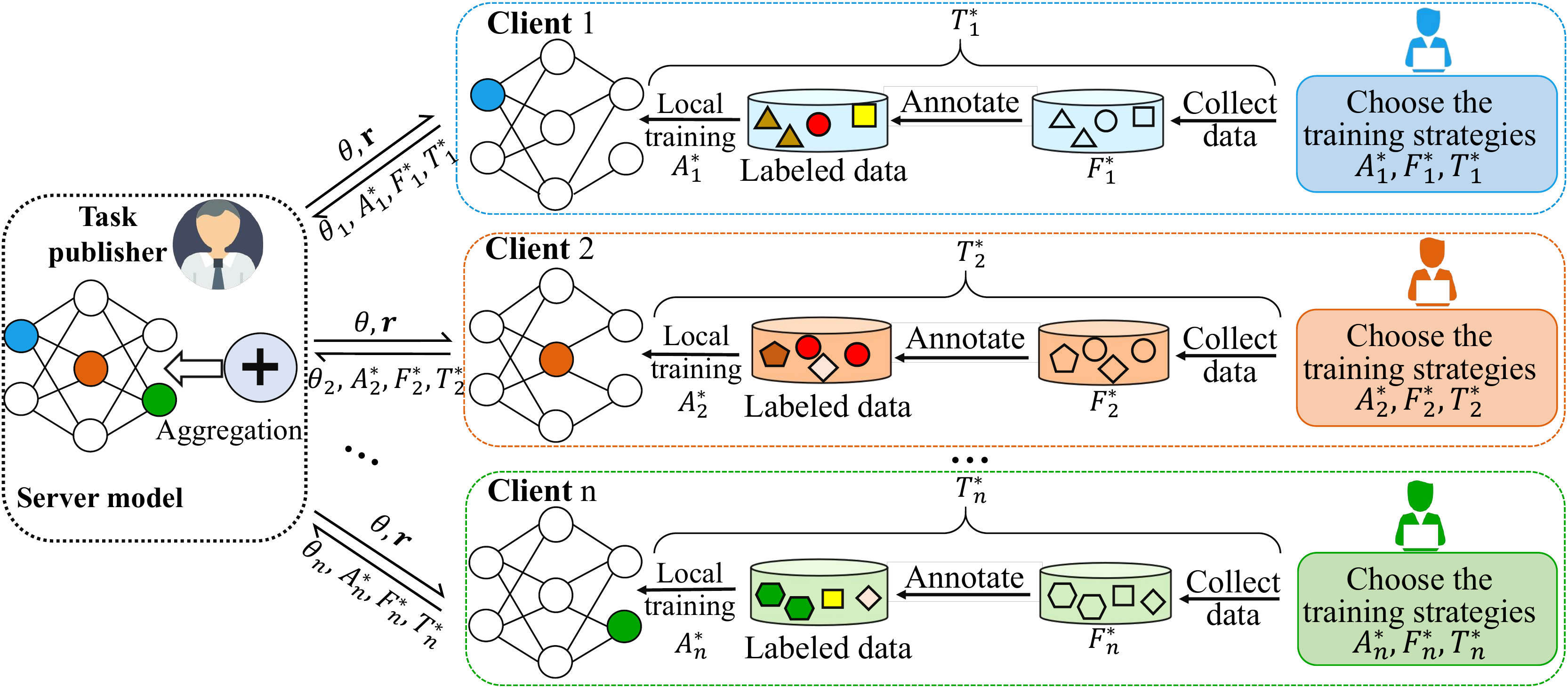}
    \caption{
    Schematic framework of iFedCrowd. The task publisher publishes a to-be-trained model $\theta$ and the requirements of training data. Then the server first announces the reward rates $\mathbf{r}$ to the participating clients. For any client $k$, it chooses the training strategy at the given reward rates: the accuracy level $A_k^*$,  data freshness $F_k^*$ and completion time $T_k^*$. Next, the client collects the required  data and completes training task for attaining the accuracy  $A_k^*$ and  the data freshness $F_k^*$ within the limited time $T_k^*$. At last, the server aggregates the received client models $\{\theta_k\}_{k=1}^n$ to obtain  the final server model and sends the rewards to clients based on their contributions.
    }
    \label{fig:framework}
\end{figure*}

Although FL has shown great advantages in privacy-preserving crowdsourcing, it still faces an open challenge that how to incentive clients to participate in the FL by contributing their computational/communication resource and data~\cite{zhan2021incentive}. Clients may be reluctant to perform local training and share their model updates without sufficient compensation. Moreover, although FL does not require participants to upload their raw data to the remote server, the malicious attackers and  the curious  server may still infer the private information of training data from the intermediate model parameters and gradients~\cite{song2017machine}. Such security risks and  potential threats aggravate the reluctance of client participation~\cite{song2020analyzing}. On the other hand, sufficient rewards can motivate them to tolerate these risks and make contributions. In addition, the clients in FL are independent, and only their owners can determine the participating strategy (i.e., when, where and how to participate in FL~\cite{zhan2021survey}). Hence, the rewards can affect the clients' decisions and training strategies, and the final model performance. Taken together, the incentive mechanism is essential for FL and crowdsourcing. 

Contemporary studies focus on the \emph{incentive mechanism} design for FL and are generally driven by clients' contribution, reputation and resource allocation~\cite{ding2020optimal, zhan2021survey}. They aim to accurately  evaluate the contributions of different data providers so that the revenue can be distributed reasonably, or motivate clients to contribute their computation power and bandwidth to achieve a fast convergence. Unfortunately, these incentive techniques for FL \emph{cannot} be directly applied to federated crowdsourcing. This is because: (i) The crowd workers in federated crowdsourcing continuously collect new data during the training process to perform model updating. Accordingly, the motivation of the incentive mechanism for federated crowdsourcing is to stimulate workers to use fresh data to update models. (ii) As the edge devices of crowd workers feature highly heterogeneous resources (e.g., computing power, bandwidth, or memory), the required time to upload model updates may vary significantly, thus leading to a long completion time of the training task. Hence, the time cost for local training and model uploading needs to be considered to achieve a faster model convergence rate. (iii) The collected training samples on workers' devices are annotated by themselves, some of which may be error-prone and noisy. Furthermore, potentially malicious workers may also submit low-quality data for quick pay. As such, it is essential to evaluate the data quality for different data providers to complete federated crowdsourcing project with high quality. (iv) In federated crowdsourcing, many data owners may not actively participate in the federated learning process, especially when the data owners are individual workers rather than enterprises. Therefore,  distributing  remuneration in a timely manner is crucial for recruiting and retaining more high-quality workers over time. On the other hand, 
the task publisher (server) aims to minimize the total reward, while each client has its own interests of maximizing the revenue that is defined by the received reward from the platform minus its cost of data collection and model training. 

To address above issues, we propose the \underline{i}ncentive-boosted \underline{Fed}erated \underline{Crowd}sourcing (\textbf{iFedCrowd}) that spurs mobile clients of the federated crowdsourcing market to actively collect local data and train client models for improving the server model. iFedCrowd formulates the above problem as a {Stackelberg game \cite{zhang2009stackelberg}} to analyze such scenario. In the  lower level of the game, iFedCrowd  distributes the revenue in terms of workers' local accuracy level, data freshness and total completion time, thereby encouraging  workers to accomplish the collaborative training task with high quality and efficiency. Meanwhile, it takes into account the cost of collecting data, computation and communication to reward workers with reasonable compensation so that they actively participate in the federated learning task. In the upper level of the game, {iFedCrowd maximizes the utility  of the task publisher that is defined by the obtained profit of the aggregated model minus the total reward paid to clients}. We derive the Nash Equilibrium that describes the steady state of the whole federated crowdsourcing system.  Figure~\ref{fig:framework} presents the schematic framework of iFedCrowd. The main contributions of our work are outlined as follows:\\
\noindent(i) We study how to motivate crowd workers to accomplish federated crowdsourcing projects in an economic way. We propose an incentive-boosted federated crowdsourcing solution (iFedCrowd) and formulate this solution as a Stackelberg game, which motivates workers to constantly collect fresh data and refine client models.\\
\noindent(ii) We derive the Nash Equilibrium of the Stackelberg game to obtain the optimal solution that maximizes the profit of the task publisher and the participating clients.\\
\noindent(iii) Extensive simulations are conducted to demonstrate that iFedCrowd can motivate workers to complete secure crowdsourcing projects with high quality and efficiency.

\section{Related Work}
Our work is closely related with the researches from two branches: privacy protection of crowd workers and incentive mechanisms in federated learning. 

As the crowdsourced data is collected by humans, the data submitted by workers involves private information and may cause serious privacy leakage~\cite{ryoo2017privacy, xu2019blockchain}. 
Differential privacy~\cite{dwork2008differential} is a widely-adopted technique to protect participants' privacy. However, employing such data perturbation techniques needs to inject strong noise into raw data or intermediate results, which severely deteriorates data accuracy. Various encryption techniques have also been applied to circumvent the exposure of private information. To name a few,  \citet{tang2020privacy} proposed a privacy-preserving task recommendation scheme with win-win incentives in crowdsourcing through developing attribute-based encryption with preparation/online encryption and outsourced decryption technologies. \citet{wu2019privacy} proposed a privacy-aware task allocation and data aggregation scheme (PTAA) that leverages bilinear pairing and  homomorphic encryption. \citet{miao2019privacy} presented a privacy-preserving truth discovery framework, which performs weighted aggregation on users' encrypted data using a homomorphic cryptosystem. Nevertheless, these encryption-based methods would bring complex computations for data processing  and analysis, and cannot defend against privacy inference attacks~\cite{wang2019towards, yuan2019priradar}. 

To relieve these disadvantages, federated learning (FL) 
allows multiple clients to collaboratively train a shared model by iteratively aggregating model updates without exposing their raw data. CrowdFL \cite{zhao2021crowdsensing} integrates FL into mobile crowdsensing and enables participants to locally process collected data via FL and only upload encrypted training models.  \citet{zhang2020light}  utilized a statistical iterative crowdsourcing algorithm to combine inference results from  different FL client models. However, these FL-enabled crowdsourcing methods follow a too optimistic assumption that crowd workers are voluntarily participating, without any returns.

The incentive mechanism is essential and crucial to FL. Since the model training operations at edge nodes will consume various resources, such as computation power, bandwidth and battery, the edge nodes would not like to get involved in this voluntary collaboration, without any compensation. Consequently, a plethora of studies  have concentrated on incentive mechanism design in FL. \citet{song2019profit} proposed the contribution index based on Shapley value to evaluate the contribution of different clients. They reconstructed the approximate models  on different combinations of training datasets through the intermediate results so as to effectively calculate the contribution index.  \citet{zeng2020fmore} presented a procurement auction incentive framework considering the multi-dimensional and dynamic edge resources. They applied the game theory to derive the optimal strategies for each client, and leveraged the expected utility to guide the parameter server to select the optimal clients to train the learning model.  \citet{lim2020hierarchical} used the contract theory to build the incentive mechanism between clients and users, and the coalitional game theory to reward the clients based on their marginal contributions. \citet{yu2020sustainable} proposed a fair incentive scheme to achieve the long-term system performance and avoid the unfairness treatment during the training process. {\citet{pandey2020crowdsourcing} incorporated FL into the crowdsourcing framework and formulated a two-stage Stackelberg game to enhance the communication efficiency. \citet{zhan2020big} introduced a game-based incentive mechanism to motivate crowd workers to maximally contribute  their local data for FL learning task.} 

However, all these studies  are inapplicable to federated crowdsourcing scenario where a worker continuously collects new data samples. In addition, they disregard the completion time of federated learning tasks, 
and the  instability of crowd worker's participation.
{Our proposed iFedCrowd utilizes the Age of Information (AoI)~\cite{li2019general} to quantify the freshness of collected data. It rewards the workers that can provide fresh data with more remuneration, thus encouraging workers to constantly  collect the suitable task data.  In addition, it takes the completion time spent on data collection and model training into account to measure the contribution of workers, so as to incentivize workers to accomplish the task at a faster pace.}

\section{Methodology}
\subsection{Problem Definition}
Let $\mathcal{W} = \{ w_1, w_2,  \cdots, w_n\}$ be the $n$ crowd workers participating in federated crowdsourcing.  Each worker $w_k \in \mathcal{W}$ collects his/her own set of tasks for annotations $\mathcal{D}_k = \{ \mathbf{x}_1^k, \mathbf{x}_2^k, \cdots, \mathbf{x}_{N_k}^k \}$, where $N_k$ denotes the number of tasks collected by worker $w_k$. Let $\mathcal{Y}_k = \{ y_1^k, y_2^k, \cdots, y_{N_k}^k \}$ be the  labels annotated by worker $w_k$ for the corresponding $N_k$ tasks. To preserve the privacy of participants, both $\mathcal{D}_k$ and $\mathcal{Y}_k$ are kept locally and not shared with the FL server or other clients.  We propose iFedCrowd to train the server model $\tilde{\theta}$ via the collaboration with $n$ client models $\{ \theta_1, \theta_2, \cdots, \theta_n \}$.

\subsection{Stackelberg Game based Incentive Mechanism}
In federated crowdsourcing, the task publisher sets up a to-be-trained model $\theta$ and the requirements for training data. Then iFedCrowd recruits crowd workers to collect suitable training data and collaboratively train the shared model.  Meanwhile, the task publisher allocates rewards to the participating clients to achieve an optimal local accuracy, and incentivizes clients for maximizing its own benefits, i.e., a well-trained model with low budget. Upon receiving rewards from the server, the rational clients will individually maximize their own profits. Such interaction scenario between the server and clients can be viewed as a Stackelberg game. {The game can be divided into two levels. In the lower level, the participating clients independently determine their strategies to solve the  local subproblem with the offered incentive. In the upper level, the task publisher decides on the reward rates for clients to build a high-quality model and maximize the utility. } 

\textit{Lower-level Subgame:} In the lower level of the game, the task publisher will firstly announce  uniform reward rates for the participating clients. Intuitively, for a higher accuracy of the client model trained over the local data, fresher collected data and less completion time, there will be an increase in the reward for the participating clients. Therefore, the revenue allocated to client $k$ is defined as follows:
\begin{equation}
    v_k = r_1 \cdot \frac{A_k}{T_k} + r_2 \cdot F_k
    \label{Allocated reward}
\end{equation}
where $A_k, F_k, T_k$ are the accuracy level of the client model $\theta_k$, the freshness of the training data, and the completion time of local training and model update, respectively. $r_1$ \textgreater ~$0$ and $r_2$ \textgreater ~$0$ are the corresponding reward rates. To compute the freshness of the collected training data, the server requires the workers to record the time at which the data was collected. Let $g_k(t)$ be the generation time of the client $k$'s most recent training sample  at time slot $t$, we can define the freshness of the data as:
\begin{equation}
    F_k = \frac{1}{t - g_k(t)}
    \label{freshness}
\end{equation}
Then rational workers will try to improve the freshness of the training data, shorten the completion time, and improve the local model's accuracy for maximizing its utility.

At the same time, training the global model (i.e., the to-be-trained model) with local data for a defined accuracy level and limited training time incurs a cost for the participants, mainly including the calculation cost and the communication cost:
\begin{equation}
    C_k = c_k^{cal} + c_k^{col} + c^{com}
    \label{Cost}
\end{equation}
where $c_k^{cal}$, $c_k^{col}$  and $c_k^{com}$ denote the calculation cost, data collection cost  and communication cost, respectively. $c_k^{cal}$ is related with the number of iterations to train the local model for attaining  the target accuracy $A_k$. Based on the relation between local iterations and model accuracy in \cite{pandey2020crowdsourcing}, we define the calculating cost for client $k$ as:
\begin{equation}
    c_k^{cal} = \gamma_k (1+A_k) \log (1 + A_k)
    \label{computing cost}
\end{equation}
where $\gamma_k$ \textgreater ~0 is a parameter choice of client $k$ that depends on the local data size and the condition number of the local subproblem. 
Hence, more iterations  result in more calculation costs on clients' devices.  

The cost of collecting training data for workers is proportional to the data freshness. To guarantee the freshness, workers should continuously collect new data. Therefore, the data collection cost for client $k$ can be defined as:
\begin{equation}
    c_k^{col} = e^{\delta_k \cdot F_k}
\end{equation}
where $\delta_k$ \textgreater~ 0 is a parameter of client $k$ that depends on the performance of sensors on workers' devices.

The communication cost $c^{com}$ is the same for all the participating clients and  is incurred when a client interacts with the server for model update. During the iterative process of the collaborative training task, let $s$ be the size of the model parameters, this total cost can be defined as $c^{com}$=$s$.

With the reward allocated to workers defined in Eq.~\eqref{Allocated reward} and the cost of participating clients  defined in Eq.~\eqref{Cost}, the \emph{client utility model} for workers can be defined as:
\begin{equation}
    u_k =r_1 \cdot \frac{A_k }{T_k} + r_2 \cdot F_k - C_k
    \label{Client utility model}
\end{equation}

\textit{Upper-level Subgame:} In the upper level of the game, the task publisher can determine the optimal reward rates $\mathbf{r}^*$ ([$r_1^* , r_2^*$]) to maximize the profit after knowing the response of workers. The utility of the task publisher can be defined by the final model performance and total completion time. As a result, the  \emph{server utility model} is defined as follows:
\begin{equation}
    U = \frac{1}{n} \sum_{k = 1}^n  (\alpha \cdot A_k + \beta \cdot F_k) - \max_{1 \leq k \leq n} T_k - \sum_{k = 1}^n (r_1 \cdot \frac{A_k}{T_k} + r_2 \cdot F_k)
\label{task_publisher_utility}
\end{equation}
where $\alpha$ \textgreater ~0 and $\beta$ \textgreater ~0 are the system parameters and $\sum_{k = 1}^n (r_1 \cdot \frac{A_k}{T_k} + r_2 \cdot F_k)$ is the total cost spent for incentivizing workers to participate in the federated learning task.

{Based on the game formulation defined above, we consider the optimal choice that maximizes the utility of both the task publisher and the participating clients. Hence, we derive the Nash Equilibrium to find the optimal solution for the two subgames in the next subsection.}

\subsection{Nash Equilibrium}

\textit{Definition 1. Nash Equilibrium. ($\mathbf{r}^*$, $\mathbf{A}^*$, $\mathbf{F}^*$, $\mathbf{T}^*$) is a Nash Equilibrium if it satisfies the following conditions:}
\begin{equation}
    U (\mathbf{r}^*, \mathbf{A}^*, \mathbf{F}^*, \mathbf{T}^*) \geq U (\mathbf{r}, \mathbf{A}^*, \mathbf{F}^*, \mathbf{T}^*)
    \label{Server eqilibrium}
\end{equation}
\begin{equation}
    u_k (\mathbf{r}^*, A_k^*, F_k^*, T_k^*) \geq u_k (\mathbf{r}^*, A_k, F_k, T_k), \forall k \in \mathcal{W}
    \label{Client eqilibrium}
\end{equation}
\textit{for any values of $\mathbf{r}$, $\mathbf{A}$, $\mathbf{F}$, and $\mathbf{T}$.}

{To study the equilibrium of the lower-level game, we derive the best response for each client.






\textit{Theorem 1. The client $k$'s best response regarding the target accuracy $A_k^*$, data freshness $F_k^*$ and completion time $T_k^*$ can be characterized as follows:}
\begin{equation}
        A_k^* = e^{h_k(T_k^{min})} - 1,
        F_k^* = \frac{1}{\delta_k} \log (\frac{r_2}{\delta_k}), 
        T_k^* = T_k^{min}  
\label{optimal_response}
\end{equation}
\textit{where $h_k(T_k^{min})$ is $\frac{r_1}{\gamma_k \cdot T_k^{min}} - 1$ and $T_k^{min}$ is the minimum time for worker $k$ to complete the data collection and model training.}

\textit{Proof.} See the supplementary file for the detailed proof.




According to \textit{Theorem 1}, the task publisher, which is the leader in the Stackelberg game, can  derive the unique Nash Equilibrium among participating clients under any given reasonable reward rates $\mathbf{r}$. Consequently, the task publisher can  maximize its utility by choosing the optimal reward rates (i.e., the equilibrium of the upper-level game). The utility model of the task publisher based on the set of best response $\mathbf{A}^*$, $\mathbf{F}^*$ and $\mathbf{T}^*$ is defined as follows: 
\begin{equation}
    U (\mathbf{r}) = \frac{1}{n} \sum_{k = 1}^n  (\alpha \cdot A_k^* + \beta \cdot F_k^*) - \max_{1 \leq k \leq n} T_k^* - \sum_{k = 1}^n (r_1 \cdot \frac{A_k^*}{T_k^*} + r_2 \cdot F_k^*)
\label{task_publisher_utility_with_response}
\end{equation}

\textit{Theorem 2. The second order derivatives of $U(\mathbf{r})$ satisfy the following conditions:}
\begin{equation}
     \frac{\partial^2 U}{\partial r_1^2} \textless ~0, ~\frac{\partial^2 U}{\partial r_2^2} \textless ~0
    \label{the conditions of the second order derivatives}
\end{equation}

\textit{Proof.} See the supplementary file for the detailed proof.





Since $\frac{\partial^2 U}{\partial r_1^2}$ \textless ~0 and $ \frac{\partial^2 U}{\partial r_2^2}$ \textless ~0, the utility of the task publisher $U(\mathbf{r})$ is a strictly concave function.  Thus it has a unique maximizer $\mathbf{r}^*$ that satisfies the following conditions:

\begin{equation}
    \begin{aligned}
    \frac{\partial U} {\partial r_1}=0, \ 
    \frac{\partial U}{\partial r_2}=0
    \end{aligned}
\label{optimal_reward_rates}
\end{equation}


Therefore, there exists a unique Nash Equilibrium of the Stackelberg game. If clients do not satisfy $A_k^*$ and $F_k^*$, the server can update the server utility model according to the actual training strategy submitted by clients, and recalculate the optimal reward rates. It will assign new reward rates in the next training round and still achieve the maximum server profit.}

\begin{algorithm}[t]
\caption {\textbf{iFedCrowd}: \underline{i}ncentive-boosted \underline{Fed}erated \underline{Crowd}sourcing}
\label{alg:Stackelberg}
\textbf{Input}: $n$ clients, local datasets  $\{\mathcal{D}_c\}_{c=1}^n$; the computation parameters $\{\gamma_k\}_{k=1}^n$; data collection parameters $\{\delta_k\}_{k=1}^n$; size of the published model $s$;  system parameters $\alpha$ and $\beta$.  \\
\textbf{Output}: Global model $\tilde{\theta}$.
\begin{algorithmic}[1] 
        \STATE \textbf{Procedure at the Central Server:}
        \FOR {all clients $k =1 \to n$ \textbf{in parallel}}
            \STATE Calculate the client $k$'s utility $u_k$ given the reward rates $r_1$ and $r_2$ via Eq.~\eqref{Client utility model}.
            \STATE Compute the optimal response including the accuracy level $A_k^*$, the data freshness $F_k^*$ and the completion time $T_k^*$ via
            Eq.~\eqref{optimal_response}.
        \ENDFOR
        \STATE Calculate the task publisher's utility via Eq.~\eqref{task_publisher_utility_with_response}.
        \STATE Determine the optimal reward rates $r_1^*$ and $r_2^*$ via Eq.~\eqref{optimal_reward_rates}, then announce them to all clients.
        \STATE Wait for all clients to complete the data collection and  model training task.
        \STATE Receive the updated client models $\{\theta_k\}_{k=1}^n$ and send the rewards to clients based on their contributions.
        \RETURN the aggregated server model $\tilde{\theta}$.
        \STATE \textbf{Procedure at Local Client $k$:}
        \STATE Receive the reward rates $r_1^*$ and $r_2^*$.
        \STATE Calculate the local utility via  Eq.~\eqref{Client utility model}.
        \STATE Choose the training strategies including $A_k^*$, $F_k^*$ and $T_k^*$ to solve the local subproblem via  Eq.~\eqref{optimal_response}.
        \STATE Collect the required data and complete training task for attaining the accuracy level $A_k^*$ and  the data freshness $F_k^*$ within the limited time $T_k^*$.
        \STATE Send the updated local model $\theta_k$ to the server.
\end{algorithmic}
\end{algorithm}

Algorithm~\ref{alg:Stackelberg} summarizes the pseudo-code of iFedCrowd. Lines 2-10 and lines 12-16 are the procedures at central server and local clients, respectively. Specifically, the process at the server consists of the following steps: compute the optimal response at the given reward rates for each client (lines 2-5), calculate the optimal reward rates and distribute them to the participating clients, then wait for them to finish the task (lines 7-9), receive the client models and return the aggregated model to the task publisher (lines 10). The process at the local client consists of the following steps: receive the announced reward rates from the server (line 12), choose the optimal data collection and model training strategies and accomplish the set goals (lines 14-15), and send back the updated local model to the server (line 16).

\section{Experiments}

\subsection{Performance Comparison with Baselines}
{To characterize and demonstrate the efficacy of the proposed incentive mechanism for federated crowdsourcing, we conduct a comparison of its performance with two baselines, namely \textbf{Random} and \textbf{MAX}. \textbf{Random} randomly selects the reward rates to incentivize the participating clients. \textbf{MAX} chooses the largest revenue rates to achieve the best response. { Since \citet{pandey2020crowdsourcing} aimed to minimize the communication budget between server and users, while \citet{zhan2020big} aimed to maximize the quantity of client training data, we do not take these two most related methods for comparison.} We fix the number of crowd workers to 10. The task publisher's utility model is defined with parameters $\alpha = 80$, $\beta = 50$. Other reasonable values for the system parameters can also be used here, which do not affect performance comparisons between these methods.}  The code of iFedCrowd is shared at www.sdu-idea.cn/codes.php?name=iFedCrowd. We implement iFedCrowd in Python 3.7 with the Mindspore deep learning framework.


To verify the effectiveness of the optimal reward rates, we evaluate the performance of baselines under different configurations, that is,  with different client parameters including $\gamma$ and $\delta$. To investigate the impact of $\gamma$, we set $\gamma$ to be uniformly distributed on $[\Gamma, \Gamma + 4](\Gamma = 1, 2, \cdots, 6 )$ and $\delta$ to be distributed on $[1, 2]$. In the similar way, we set $\delta$ to be uniformly distributed on $[\Delta, \Delta + 1] (\Delta = 0, 1, \cdots, 5)$ with the  uniformly distributed $\gamma$ on $[1, 5]$ to analyze the impact of $\delta$. We independently run each method ten times and record the average performance. 

\begin{figure}[tp]
\centering
\includegraphics[width=8cm,height=3.5cm]{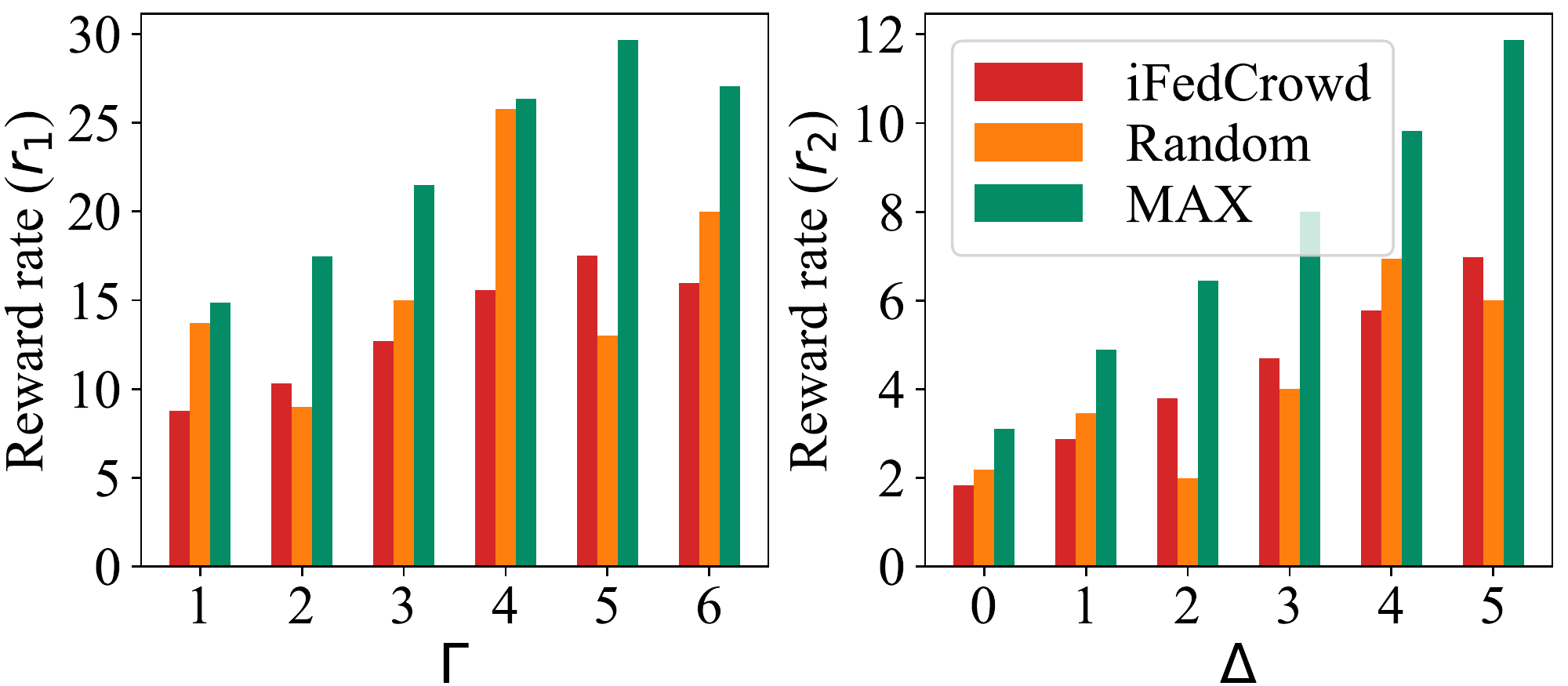}
\vspace{-0.9em}
\caption{Reward rates ($r_1$ and $r_2$) vs. client parameters ($\gamma \in [\Gamma, \Gamma+4]$ and $\delta \in [\Delta, \Delta+1]$).}
\label{r_1_r_2}
\end{figure}

Figure \ref{r_1_r_2} reports the reward rates  of iFedCrowd and the baselines under different configurations. We have the following observations: (i) All the three methods have a significant increase in reward rates as the computing cost parameter $\gamma$ and the data collection cost parameter $\delta$ are enlarged. The reason is that attaining the same accuracy level and collecting the new training samples consume much more budget, and thus the participating clients exert more incentive to compensate for their cost. (ii) \textbf{MAX} always allocates the largest reward rates  to participating clients since it aims to encourage workers to collect more fresh data and achieve the highest local accuracy as much as possible. However, {this strategy also wastes much more budget and allows the task publisher to reap very few profits}. (iii) \textbf{Random} randomly selects reward rates for workers regardless of the computing cost, data collection cost and the communication cost of clients. {Therefore, it may allocate inadequate compensation to workers and cannot maintain a stable profit for the task publisher}. (iv) Our iFedCrowd achieves a more significant improvement in choosing reward rates than baselines. This is because iFedCrowd considers the task publisher's profit and the cost spent for incentivizing workers to make contributions. Then it determines the optimal reward rates to maximize the utility of the participating clients and the task publisher. As a result, iFedCrowd can attract workers to make significant contributions with small reward rates.

\begin{figure}[tp]
\centering
\includegraphics[width=8cm,height=3.5cm]{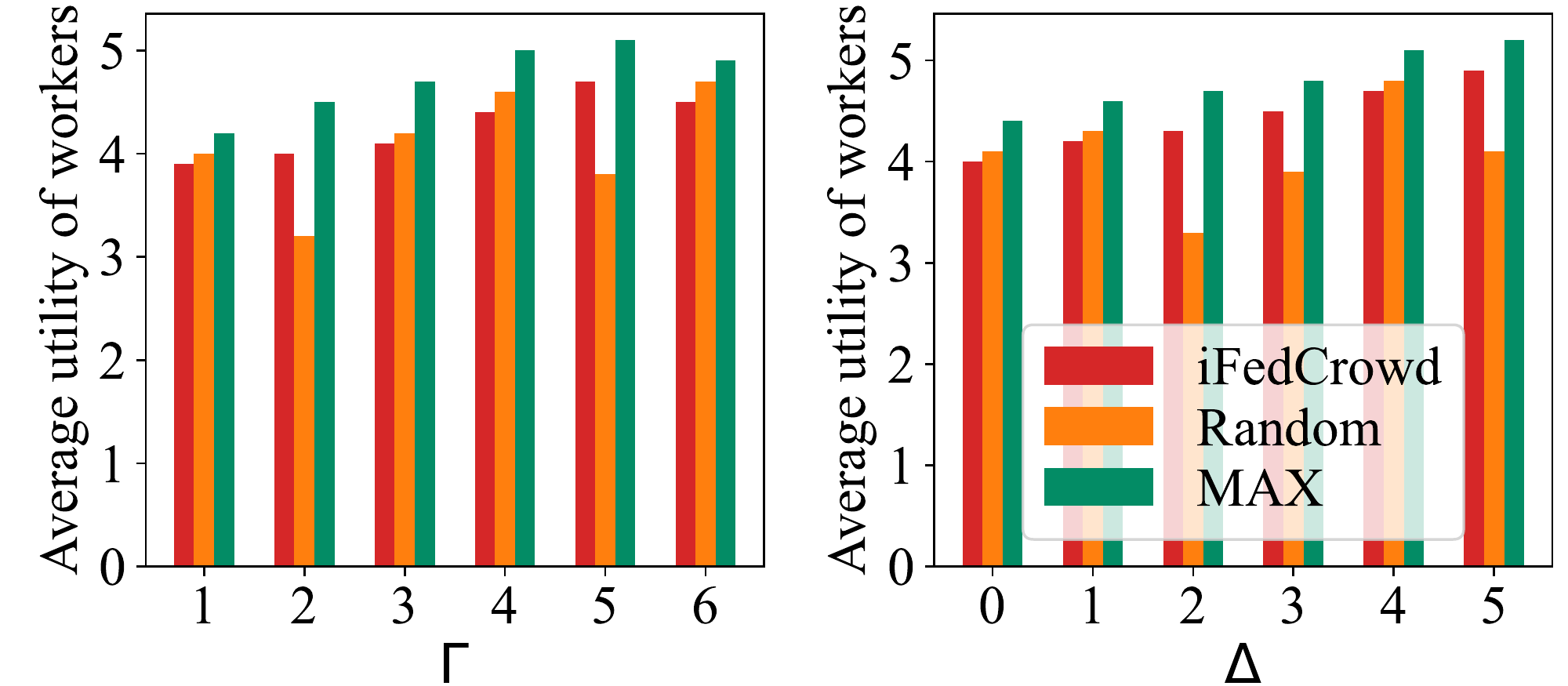}
\vspace{-0.9em}
\caption{Average utility of workers vs. client parameters ($\gamma \in [\Gamma, \Gamma+4]$ and $\delta \in [\Delta, \Delta+1]$).}
\label{client_utility}
\end{figure}

To evaluate the attractiveness of iFedCrowd to crowd workers, we plot the average utility of workers with different client parameters in Figure~\ref{client_utility}. The results under different configurations give similar observations, and iFedCrowd achieves the competitive client utility than other baselines with the increasing client parameters. In addition, we have the following important observations: (i) \textbf{MAX} has a better performance than \textbf{Random} in most cases, which offers much more reward to stimulate workers making greater contributions. Nevertheless, iFedCrowd  achieves the competitive utility over \textbf{MAX}, which wastes a lot of unnecessary budget to achieve the same client benefits as iFedCrowd. (ii) \textbf{Random} allocates much less revenue to the participating workers than \textbf{MAX}. In addition, the paid remuneration to workers is not stable across the different scenarios, thus aggravating the reluctance of clients' participation. (iii) In federated crowdsourcing, the individual workers may not actively participate in the published federated learning tasks. Accordingly, our proposed iFedCrowd leverages the client utility model to quantify the benefits of workers and chooses the optimal reward rates to engage the data owners. As a result, it maintains a more profitable and stable  federated crowdsourcing market, thus can attract more data owners to actively contribute their resources.

\begin{figure}[tp]
\centering
\includegraphics[width=8cm,height=3.5cm]{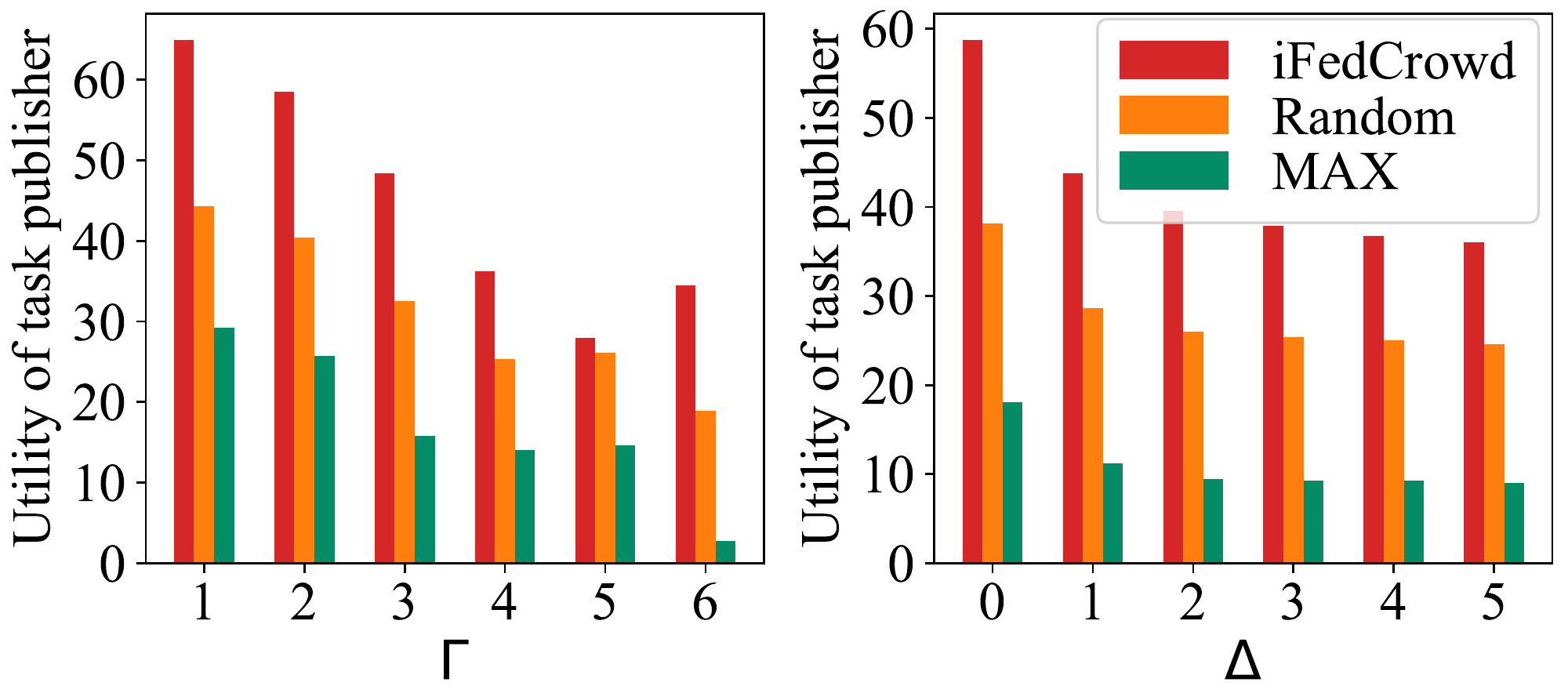}
\vspace{-0.9em}
\caption{Utility of task publisher vs. client parameters ($\gamma \in [\Gamma, \Gamma+4]$ and $\delta \in [\Delta, \Delta+1]$).}
\label{Task_publisher_utility}
\end{figure}

The bar charts in Figure~\ref{Task_publisher_utility} further present the task publisher's profit for each method with different client parameters. As expected, the larger computing and data collection costs lead to a lower utility of the task publisher.  In addition, we have the following observations: (i) \textbf{MAX} usually performs the worst as it distributes rewards to clients that greatly outweigh the benefits of clients' contributions, thus consuming much more cost and resulting in less profit of the task publisher. (ii) \textbf{Random} assigns much less reward to the participating clients since it randomly chooses the reward rates. Hence, it allows the task publisher to obtain more profit than that of \textbf{MAX}. However, it does not take into account the optimization of task publisher's utility given the response of participating clients and just randomly determines the reward rates of the utility model. Therefore, it is outperformed by iFedCrowd. (iii) iFedCrowd clearly outperforms the compared baselines in different scenarios. This is because iFedCrowd utilizes the server utility model to select the optimal reward rates to maximize the profit of the task publisher.  Therefore, it enables the task publisher to efficiently obtain a high-quality server model with low budget. 

{
\subsection{Experiment with real crowdsourcing project}
We used a real-world dataset called FitRec~\cite{ni2019modeling} for experiments. FitRec dataset is collected from the sport website Endomondo and includes multiple sources of sequential sensor data generated on mobile devices, such as heart rate, speed, GPS, sport type and user gender.
Following~\cite{ni2019modeling}, we re-sample the sequential data in 10-second intervals, and further generate two derived sequences: distance and speed. 
We randomly select 50 users as crowd workers to participate in the federated crowdsourcing. A single layer LSTM followed by a fully connected layer is used as the backbone for the training model.   We use the canoncial RMSE (Root Mean Squared Error) as the evaluation metric to quantify the performance on the prediction tasks.

\begin{table}[h!tbp]
\centering
\caption{Experimental results on real-world FitRec dataset}
\begin{tabular}{c|c|c|c}
\hline
  &  {\bf Random} & {\bf MAX} & {\bf iFedCrowd} \\
\hline
$r_1$ & 17.189 & 23.957 & 12.468 \\
$r_2$    & 1.568 & 7.497 & 4.562 \\
Worker utility  & 4.713 & 6.216 & 4.165 \\
Server utility & 37.197 & 14.223 & 52.518 \\
RMSE & 4.613 & 3.634 & 3.057 \\
\hline
\end{tabular}
\label{tab:Results on Fitrec}
\end{table}

As shown in Table~\ref{tab:Results on Fitrec},  {\bf MAX} allocates the largest reward rates to the users, as expected. As such, {\bf MAX} offers the highest worker utility among the three methods. However, it just obtains a  server model that is comparable in performance with iFedCrowd. In other words, it wastes much more budget and results in the lowest server utility.  iFedCrowd achieves the competitive worker utility than other baselines and clearly outperforms the compared baselines in terms of server utility. This is because iFedCrowd estimates the training strategy for each client in advance. Then it utilizes the server utility model to select the optimal reward rates to achieve the best performing model without wasting budget. As a result, it enables the task publisher to gain the maximum profit.
}

\subsection{Impact of Number of Workers}
\begin{figure}[tp]
\centering
\includegraphics[width=8cm,height=3.5cm]{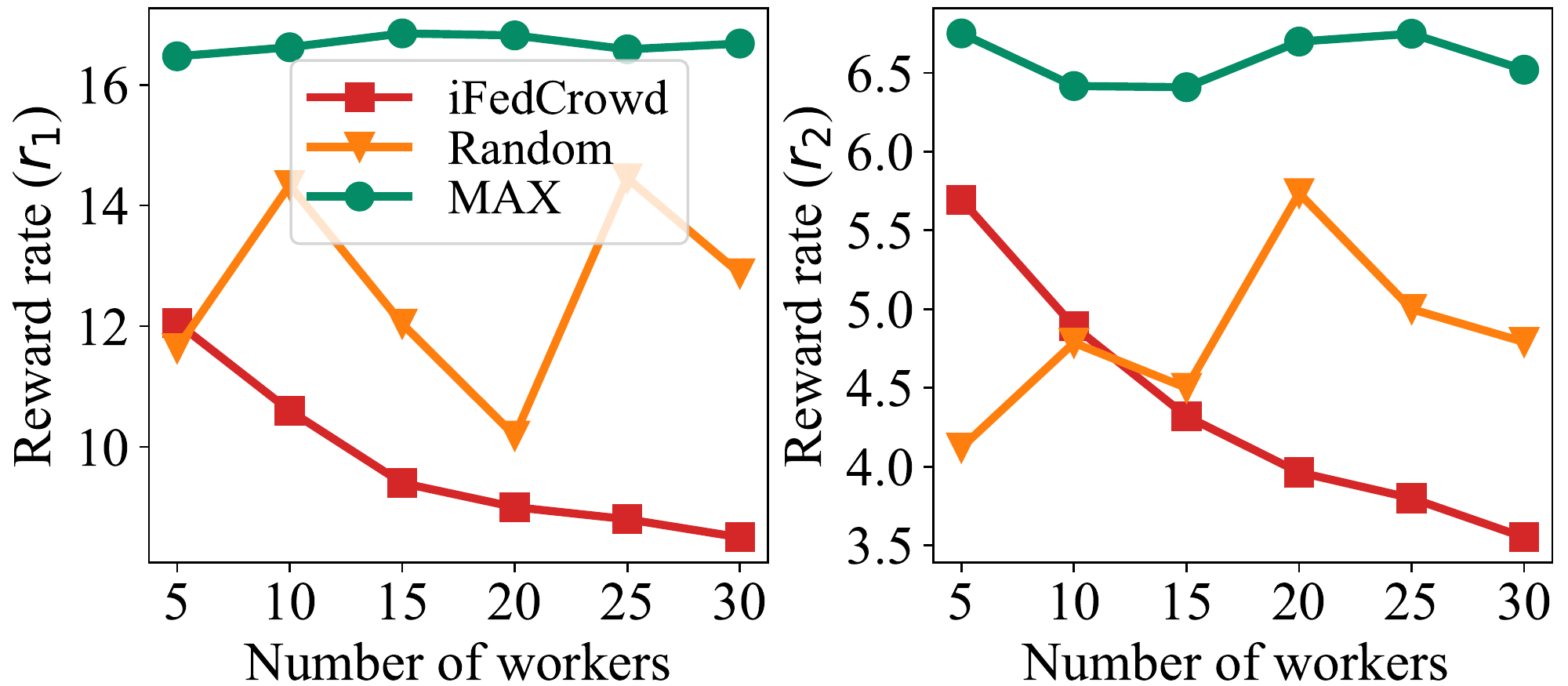}
\vspace{-0.9em}
\caption{Reward rates ($r_1$ and $r_2$) vs. number of workers.}
\label{rates_worker_number}
\end{figure}

\begin{figure}[tp]
\centering
\includegraphics[width=8cm,height=3.5cm]{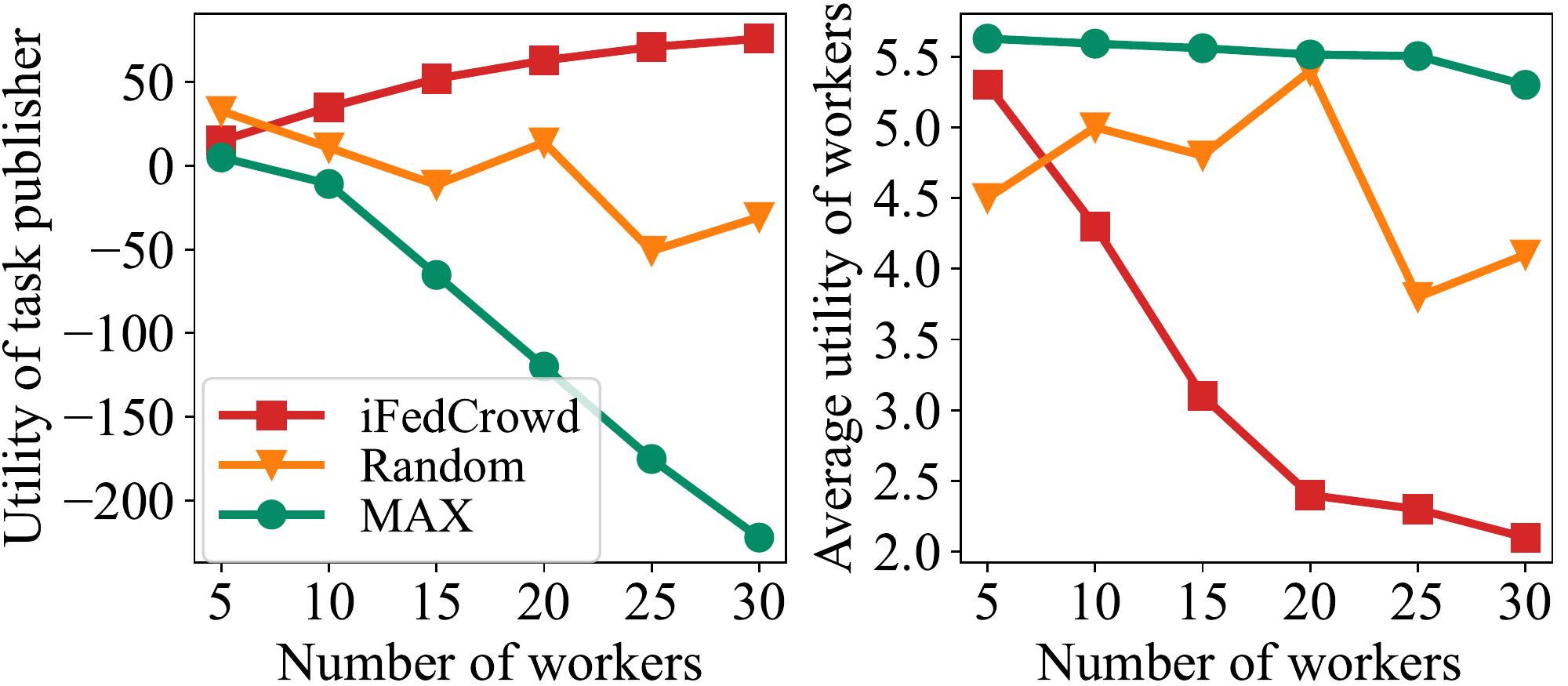}
\vspace{-0.9em}
\caption{The utility of task publisher and workers vs. number of workers.}
\label{utility_worker_number}
\end{figure}

To more explicitly evaluate iFedCrowd, we plot the reward rates and the utility of task publisher and clients with different number of participating workers in Figure~\ref{rates_worker_number} and Figure ~\ref{utility_worker_number}.
{The experimental results are relatively stable when the number of workers exceeds 30,} so 
we vary the number of workers from 5 to 30. 
{The conclusions under different client parameters are similar, so we use a reasonable and fixed range for different parameters, namely 
$\gamma \sim{U(3,5)}$ and $\delta \sim{U(2,4)}$}.

Figure~\ref{rates_worker_number} reports the impact of the number of client workers on the reward rates. We observe that: (i) \textbf{MAX} always releases the highest and inflexible  reward rates to participating clients as the number of workers increases. It aims to maximize each client's participation level, no matter how much the clients cost. Although this mechanism enhances the attractiveness to crowd workers, the obtained benefits  are not proportional to its payment, thus severely cutting down the profit of the task publisher.  (ii) In \textbf{Random} mechanism, the server determines the reward rates to workers randomly regardless of changes in the number of workers. Therefore,  the cost of \textbf{Random} is relatively low compared to \textbf{MAX}. But potential and inadequate incentives will prevent workers from contributing their data and computation resources. Furthermore, the uncertainty of the allocated revenue can not attract more workers to participate in the federated learning task. (iii) As the number of  participating workers increases, the reward rates of iFedCrowd decrease. This is because iFedCrowd considers the remuneration paid to the workers, and it chooses the optimal reward rates to maintain a high server profit when more workers joining the federated learning task. Hence, iFedCrowd  gradually reduces the reward rates to maximize the benefits for the task publisher.

In Figure~\ref{utility_worker_number}, we display the changing utility of task publisher and workers with an increasing number of workers. We notice that: (i) The average utility of workers in  \textbf{MAX} is much more inflexible and higher that of \textbf{Random} and iFedCrowd, which is in congruence with the previous analysis. Hence, the wasted budget allocated to participating clients leads to the lowest utility of task publisher, as more and more clients participate in the federated learning task. (ii)  In contrast to \textbf{MAX} and \textbf{Random}, iFedCrowd obtains an increasing utility of task publisher when the number of workers increases. This is because iFedCrowd selects the optimal reward rates according to the set of workers' response. Then it reduces the allocated reward to maximize the utility of the task publisher when more workers submit their response to the server. At the same time, this mechanism will also cause a decline in workers' utility. In addition, more workers will also lead to more competition among participating workers. Therefore, each worker will obtain less reward from the task publisher as more and more workers involved.

\section{Conclusion}
We presented an incentive mechanism iFedCrowd to complete secure crowdsourcing projects with quality and efficiency. iFedCrowd aims to jointly maximize the utility of the participating clients and the crowdsourcing platform, and it defines the Stackelberg game  to model the competition between clients and platform. We derive the best response solution and the existence of Nash Equilibrium of this game. Extensive experiments confirm the efficacy of iFedCrowd.

\section{Acknowledge}
This work is supported by National Key Research and Development Program of China (No. 2022YFC3502101), NSFC (No. 62031003 and 61872300) and CAAI-Huawei MindSpore Open Fund.



\clearpage

\twocolumn[
\begin{@twocolumnfalse}
	\section*{\centering{Incentive-boosted Federated Crowdsourcing \\
Supplementary file \\[25pt]}}
\end{@twocolumnfalse}
]

\subsection{Proof for Theorem 1}

\textit{Theorem 1. The client $k$'s best response regarding the target accuracy $A_k^*$, data freshness $F_k^*$ and completion time $T_k^*$ can be characterized as follows:}
\begin{equation}
        A_k^* = e^{h_k(T_k^{min})} - 1,
        F_k^* = \frac{1}{\delta_k} \log (\frac{r_2}{\delta_k}), 
        T_k^* = T_k^{min}  
\label{optimal_response}
\end{equation}
\textit{where $h_k(T_k^{min})$ denotes $\frac{r_1}{\gamma_k \cdot T_k^{min}} - 1$ and $T_k^{min}$ is the minimum time for the worker $k$ to complete the data collection and model training.}

\textit{Proof.} The first order derivatives of client utility model $u_k (r, A_k, F_k, T_k)$ with respect to $A_k$, $F_k$ and $T_k$ are as follows:
\begin{equation}
\begin{aligned}
    \frac{\partial u_k(r, A_k, F_k, T_k)}{\partial A_k} &=  - \gamma_k \log (A_k + 1) - \gamma_k + \frac{r_1}{T_k}  \\
    \frac{\partial u_k(r, A_k, F_k, T_k)}{\partial F_k} &= - \delta_k e^{\delta_k \cdot F_k} + r_2  \\
    \frac{\partial u_k(r, A_k, F_k, T_k)}{\partial T_k} &= - \frac{r_1 \cdot A_k}{T_k^2}\\
\end{aligned}
\label{first_order_of_client_utility_model}
\end{equation}

As $\frac{\partial u_k(r, A_k, F_k, T_k)}{\partial T_k}$ \textless ~0, $u_k(r, A_k, F_k, T_k)$ is a function that is  strictly monotone decreasing relative to the completion time $T_k$. As a result, a rational worker $k$ will try to complete the data collection and model training in the shortest time $T_k^{min}$ she/he can reach, namely $T_k^* = T_k^{min}$.

Setting $\frac{\partial u_k(r, A_k, F_k, T_k)}{\partial A_k} = 0$, we have
\begin{equation}
    \begin{aligned}
       &\frac{\partial u_k(r, A_k, F_k, T_k)}{\partial A_k} = 0 \\
       &\Leftrightarrow  \log (A_k + 1)   = \frac{r_1}{\gamma_k \cdot T_k} - 1 \\
    \end{aligned}
\end{equation}

Let $h_k(T_k)$ denotes $\frac{r_1}{\gamma_k \cdot T_k} - 1$,  we can derive
\begin{equation}
    A_k^* = e^{h_k(T_k^{min})} - 1
\end{equation}

Setting $\frac{\partial u_k(r, A_k, F_k, T_k)}{\partial F_k} = 0$, we have
\begin{equation}
    F_k^* = \frac{1}{\delta_k} \log (\frac{r_2}{\delta_k})
\end{equation}

We can therefore characterize the client $k$'s best response regarding the target accuracy $A_k^*$, data freshness $F_k^*$ and completion time $T_k^*$ as follows:
\begin{equation}
        A_k^* = e^{h_k(T_k^{min})} - 1,
        F_k^* = \frac{1}{\delta_k} \log (\frac{r_2}{\delta_k}), 
        T_k^* = T_k^{min}  
\label{optimal_response}
\end{equation}

Since $0 \textless A_k^* \textless 1$ and $F_k^* \geq 0$,  we can also derive the reasonable range for the reward rates as follows:
\begin{equation}
    \begin{aligned}
    &0 \textless A_k^* =  e^{h_k(T_k^{min})} - 1 \textless 1 \\
    &\Leftrightarrow 0 \textless h_k(T_k^{min}) \textless \log 2 \\
    &\Leftrightarrow \gamma_k \cdot T_k^{min} \textless r_1 \textless (1 + \log 2 ) \gamma_k \cdot T_k^{min} \\
    &F_k^* = \frac{1}{\delta_k} \log (\frac{r_2}{\delta_k}) \geq 0, \ r_2 \geq \delta_k \\
    \end{aligned}
\label{the_range_of_the rates}
\end{equation}

\subsection{Proof for Theorem 2}

\textit{Theorem 2. The second order derivatives of $U(\mathbf{r})$ satisfy the following conditions:}
\begin{equation}
     \frac{\partial^2 U}{\partial r_1^2} \textless ~0, ~\frac{\partial^2 U}{\partial r_2^2} \textless ~0
    \label{the conditions of the second order derivatives}
\end{equation}

\textit{Proof.} The utility model of the task publisher based on the set of best response $\mathbf{A}^*$, $\mathbf{F}^*$ and $\mathbf{T}^*$ is defined as follows: 
\begin{equation}
    U (\mathbf{r}) = \frac{1}{n} \sum_{k = 1}^n  (\alpha \cdot A_k^* + \beta \cdot F_k^*) - \max_{1 \leq k \leq n} T_k^* - \sum_{k = 1}^n (r_1 \cdot \frac{A_k^*}{T_k^*} + r_2 \cdot F_k^*)
\label{task_publisher_utility_with_response}
\end{equation}

Then the first order derivative of $U(\mathbf{r})$ is 
\begin{equation}
    \frac{\partial U} {\partial r_1} = \frac{1}{n} \sum_{k = 1}^n (\alpha \cdot \frac{\partial A_k^*}{\partial r_1})  - \sum_{k = 1}^n  \frac{1}{T_k^{min}} (r_1 \cdot \frac{\partial A_k^*}{\partial r_1} + A_k^*)
\end{equation}

\begin{equation}
    \frac{\partial U}{\partial r_2} = \frac{1}{n} \sum_{k=1}^n (\beta \cdot \frac{\partial F_k^*}{\partial r_2}) - \sum_{k = 1}^n ( r_2 \cdot \frac{\partial F_k^*}{\partial r_2} + F_k^*)
\end{equation}

Hence, the second order derivative of $U(\mathbf{r})$ is 
\begin{equation}
\begin{aligned}
    \frac{\partial^2 U}{\partial r_1^2} &=  \frac{1}{n} \sum_{k=1}^n (\alpha \cdot \frac{\partial^2 A_k^*}{\partial r_1^2}) - \sum_{k=1}^n  \frac{1}{T_k^{min}} (r_1 \cdot \frac{\partial^2 A_k^*}{\partial r_1^2} + \frac{2 \partial A_K^*}{\partial r_1}) \\
    &= - \sum_{k=1}^n (\frac{n r_1 + \alpha T_k^{min} + 2\gamma_k  n  T_k^{min}}{n \cdot \gamma_k^2 \cdot {T_k^{min}}^3} e^{\frac{r_1}{\gamma_k \cdot T_k^{min}} -1} ) \textless 0
\end{aligned}
\end{equation}

\begin{equation}
\begin{aligned}
    \frac{\partial^2 U}{\partial r_2^2} &= \frac{1}{n} \sum_{k=1}^n (\beta \cdot \frac{\partial^2 F_k^*}{\partial r_2^2}) - \sum_{k=1}^n (r_2 \cdot \frac{\partial^2 F_k^*}{\partial r_2^2} + \frac{2 \partial F_k^*}{\partial r_2}) \\
    &= \sum_{k=1}^n (-\frac{1}{\delta_k \cdot r_2} - \frac{\beta}{n \cdot \delta_k \cdot r_2^2})  \textless 0
\end{aligned}
\end{equation}

\end{document}